\documentclass[english,prl,final,twocolumn, superscriptaddress]{revtex4-1}

\usepackage[T1]{fontenc}
\usepackage[latin9]{inputenc}
\usepackage{color}
\usepackage{babel}
\usepackage{textcomp}
\usepackage{amsmath}
\usepackage{amssymb}
\usepackage{stackrel}
\usepackage{graphicx}
\usepackage[unicode=true,pdfusetitle,
 bookmarks=true,bookmarksnumbered=false,bookmarksopen=false,
 breaklinks=false,pdfborder={0 0 1},backref=false,colorlinks=true]
 {hyperref}
\begin{document}

\title{Non-invasive quantitative imaging of selective microstructure-sizes
with magnetic resonance}

\author{Milena Capiglioni}

\affiliation{Centro Atómico Bariloche, CONICET, CNEA, S. C. de Bariloche, 8400,
Argentina}

\affiliation{Instituto Balseiro, CNEA, Universidad Nacional de Cuyo, S. C. de
Bariloche, 8400, Argentina}

\affiliation{Support Center for Advanced Neuroimaging (SCAN), Institute of Diagnostic
and Interventional Neuroradiology, University of Bern, Bern, 3010
, Switzerland}

\author{Analia Zwick}
\email{analia.zwick@cab.cnea.gov.ar}

\affiliation{Centro Atómico Bariloche, CONICET, CNEA, S. C. de Bariloche, 8400,
Argentina}

\affiliation{Departamento de Física Médica, Instituto de Nanociencia y Nanotecnologia,
CNEA, CONICET, S. C. de Bariloche, 8400, Argentina}

\author{Pablo Jiménez}

\affiliation{Centro Atómico Bariloche, CONICET, CNEA, S. C. de Bariloche, 8400,
Argentina}

\affiliation{Instituto Balseiro, CNEA, Universidad Nacional de Cuyo, S. C. de
Bariloche, 8400, Argentina}

\author{Gonzalo A. \'Alvarez}
\email{gonzalo.alvarez@cab.cnea.gov.ar}

\affiliation{Centro Atómico Bariloche, CONICET, CNEA, S. C. de Bariloche, 8400,
Argentina}

\affiliation{Instituto Balseiro, CNEA, Universidad Nacional de Cuyo, S. C. de
Bariloche, 8400, Argentina}

\affiliation{Departamento de Física Médica, Instituto de Nanociencia y Nanotecnologia,
CNEA, CONICET, S. C. de Bariloche, 8400, Argentina}
\begin{abstract}
\textbf{Extracting reliable and quantitative microstructure information
of living tissue by non-invasive imaging is an outstanding challenge
for understanding disease mechanisms and allowing early stage diagnosis
of pathologies. Magnetic Resonance Imaging is the favorite technique
to pursue this goal, but still provides resolution of sizes much larger
than the relevant microstructure details on in-vivo studies. Monitoring
molecular diffusion within tissues, is a promising mechanism to overcome
the resolution limits. However, obtaining detailed microstructure
information requires the acquisition of tens of images imposing long
measurement times and results to be impractical for in-vivo studies.
As a step towards solving this outstanding problem, we here report
on a method that only requires two measurements and its proof-of-principle
experiments to produce images of selective microstructure sizes by
suitable dynamical control of nuclear spins with magnetic field gradients.
We design microstructure-size filters with spin-echo sequences that
exploit magnetization ``decay-shifts'' rather than the commonly
used decay-rates. The outcomes of this approach are quantitative images
that can be performed with current technologies, and advance towards
unravelling a wealth of diagnostic information based on microstructure
parameters that define the composition of biological tissues.}
\end{abstract}
\maketitle
The development of nanosized sensors with novel quantum technologies
is aiming at nanoscale imaging of biological tissues for unveiling
the biophysics of pathologies at such relevant scales \citep{Staudacher2015,Wangeaau8038,Barry14133,glenn2015single}.
Such imaging proposals are still based on invasive techniques. By
contrast, Magnetic Resonance Imaging (MRI) has proven to be an excellent
tool for acquiring non-invasive images, being applied on a daily basis
for clinical diagnosis. However, the weak sensitivity for detecting
the nuclear spins inherent to the biological tissues, typically limits
the spatial resolution of in-vivo MRI to hundred of micrometers in
pre-clinical scanners and to millimeters in clinical systems. This
limitation imposes a challenge for existing methods to early detect
diseases that produce changes at the cellular level \citep{Padhani2009,Drago2011,White2013,White2014,onbehalfoftheMAGNIMSstudygroup2015}.
Detecting these kind of pathologies in a development stage based on
quantitative imaging of tissue microstructure parameters, will allow
MRI to advance towards a new early diagnostic paradigm \citep{LeBihan2003,White2013,White2014,Xu2014,onbehalfoftheMAGNIMSstudygroup2015,Grussu2017,Alexander2019}.

Diffusion Weighted MRI (DWI) is a promising tool to probe microstructure
information based on monitoring the dephasing of the nuclear spin
precession due to Brownian molecular motion \citep{LeBihan2003,Grebenkov2007,Callaghan2011}.
The diffusion dynamics of molecules depends on tissue properties such
as cell sizes, density, and other morphological features. A strong
magnetic field gradient is applied to sense the microscopic motion
of spins so that the the precession frequency depends on their instantaneous
position. In addition, modulating the gradient strength as a function
of time allows to probe the time dependent diffusion process of the
molecules within tissues \citep{Stepisnik1993,Callaghan2011,Alvarez2013a,Shemesh2013}.
These dynamical control techniques are based on the Hahn spin-echo
concept \citep{Hahn1950} and its generalization to multiple echoes
\citep{Carr1954}. Within DWI they are called modulated gradient
spin-echo (MGSE) sequences, where the phase accumulated by the spin's
precession is refocused at given times allowing to infer microscopic
parameters in tissues and porous media \citep{Stepisnik1993,Shemesh2013,Drobnjak2016,Nilsson2017,Novikov2019}.
However, obtaining detailed microstructure information is still challenging
on in-vivo studies, as it requires tens of images demanding about
an hour of acquisition time \citep{Assaf2008,Alexander2010,Xu2014,Shemesh2015,Alexander2019,Novikov2019}.

The decay of the nuclear spin signal under MGSE sequences is typically
characterized by a decay-rate \citep{Grebenkov2007,Callaghan2011}.
Here, we report that these decaying signals manifest a ``decay-shift''
that can be exploited to selectively probe microstructure-sizes. We
develop size-filters based on the Non-uniform Oscillating Gradient
Spin-Echo (NOGSE) concept that contrast the signal generated by two
spin-echo sequences \citep{Alvarez2013a,Shemesh2013}. The method
probes different diffusion time scales while factoring out other relaxation
mechanisms induced by gradient-modulation imperfections and $T_{2}$
effects. We exploit the NOGSE modulations to selectively probe the
spin-echo decay-shift for producing quantitative images based on contrast
intensities that reflect the probability of finding a specific microstructure-size.
The present approach requires only two measurements, therefore significantly
reducing the acquisition time compared to state-of-the-art methods
that typically require tens of measurements to obtain quantitative
information on microstructure-sizes based on fitting parameters \citep{Assaf2008,Alexander2010,Ong2010,Xu2014,Shemesh2015,Alexander2019,Novikov2019}.
We analytically and experimentally demonstrate that these microstructure-size
filters can be implemented with current technologies presenting a
novel mechanism for quantitative and precision imaging diagnostic
tools.

\begin{figure*}
\begin{centering}
\includegraphics[width=1\textwidth]{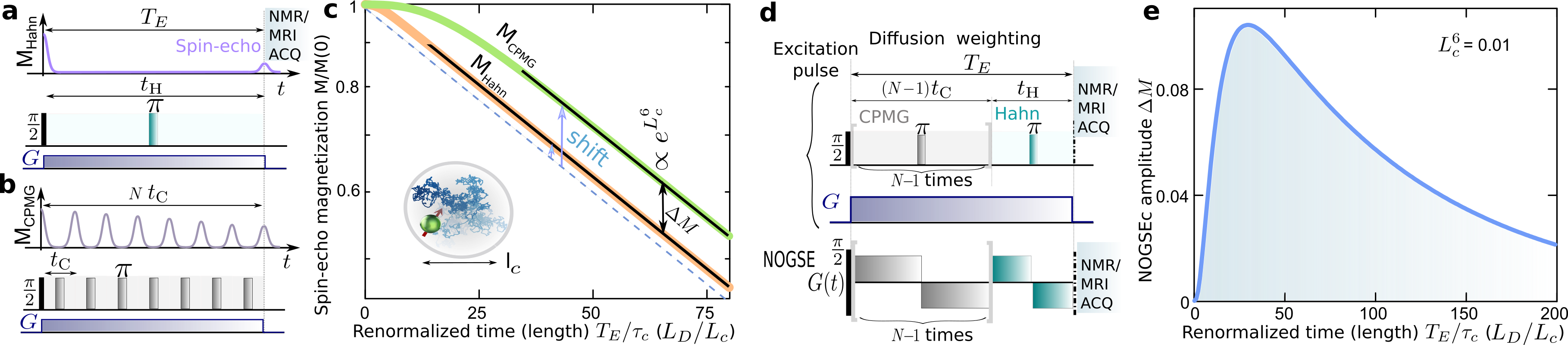}
\par\end{centering}
\caption{\textbf{Spin-echo decay-shift as a paradigm for probing microstructure
sizes. }(\textbf{a},\textbf{b}) Magnetic resonance spin-echo sequences
for DWI. An initial $\frac{\pi}{2}$-excitation pulse is followed
by a constant magnetic field gradient $G$. During the evolution time
$T_{E}$, equidistant $\pi$ rf-pulses modulate the effective gradient
$G(t)$, switching its sign for refocusing the spins dephasing induced
by the diffusion process to form the spin-echoes. (\textbf{a}) Hahn
sequence with a refocusing period $t_{H}$ ($N\!=\!1$). (\textbf{b})
CPMG sequence with $N\!>\!1$ refocusing periods of duration $t_{C}$.
After $T_{E}$ the remaining signal is measured, possibly using MRI
acquisition encoding. \textbf{(c)} Spin-echo decay for the Hahn ($N\!=\!1$,
orange line) and the CPMG ($N\!=\!8$, green line) gradient modulation
as a function of the normalized evolution time $T_{E}/\tau_{c}$,
where $\tau_{c}$ is the correlation time of the restricted diffusion
process. For both sequences, when the refocusing period is $T_{E}/N\gg\tau_{c}$,
the decaying signal \textrm{$\propto\exp\left(-\gamma^{2}G^{2}D_{0}\tau_{c}^{2}\,T_{E}\right)$}
(black solid-lines) has a constant decay-rate independent of $N$.
However, the curves have a decay-shift \textrm{$\propto\exp\left(\gamma^{2}G^{2}D_{0}\tau_{c}^{3}(1+2N)\right)$}
independent of time but depending on $N$ (marked with arrows) with
respect to the dashed-line that gives \textrm{$\exp\left(-\gamma^{2}G^{2}D_{0}\tau_{c}^{2}\,T_{E}\right)$}.
The contrast $\Delta M$ between the CPMG and Hahn decay is highlighted
in the plot with double-arrow. The inset shows a scheme for molecules
undergoing restricted Brownian motion. The diffusion restriction length
$l_{c}=\sqrt{D_{0}\tau_{c}}$ is related to the compartment-size.
(\textbf{d}) A CPMG sequence of $N\!-\!1$ equidistant refocusing
periods $t_{C}$ is concatenated with a Hahn sequence of refocusing
period $t_{H}$ \citep{Alvarez2013a}. The effective modulated gradient
$G(t)$ is shown at the bottom, and it composes the Non-uniform Oscillating-Gradient
Spin-Echo (NOGSE) sequence that can be generated by directly modulating
the gradient strength \citep{Shemesh2013}. (\textbf{e}) NOGSE contrast
(NOGSEc) $\Delta M$ as a function of the normalized evolution time
$\frac{T_{E}}{\tau_{c}}$ for \textrm{$L_{c}^{6}\!=\!\gamma^{2}G^{2}D_{0}\tau_{c}^{3}=0.01$.
}This value is representative of white-matter tissue considering $D_{0}\!=\!0.7\,\mu\mathrm{m^{2}/ms}$
and $\tau_{c}\!=\!1.5\,\mathrm{ms}$ with $G\!=\!240\,\mathrm{mT/m}$
and $N\!=\!8$. \label{fig:(a)-SDR-sequence}}
\end{figure*}

\textbf{MRI of molecular-diffusion\label{sec:Background-=000026-Theory}.}
The nuclear spins $S\!=\!\frac{1}{2}$ in molecules intrinsic to biological
tissues, mainly from water's protons, are typically observed on in-vivo
MRI. They interact with an external uniform magnetic-field $B_{0}$\textrm{$\hat{z}$}
and a magnetic field gradient $G\hat{r}$ applied along the direction
\textrm{$\hat{r}$} for spatial encoding. In a frame rotating at the
resonance frequency $\gamma B_{0}$, the precession frequency $\omega(t)=\gamma Gr(t)$
fluctuates reflecting the random motion of the molecular diffusion
process \citep{Grebenkov2007,Callaghan2011}. Here, $r(t)$ is the
instantaneous position of the nuclear spin along the gradient direction
and $\gamma$ is the gyromagnetic ratio of the nucleus. The effective
gradient strength may vary along time as $G(t)$, either by applying
$\pi$-pulses if the gradient is constant or directly by modulating
the gradient's sign and amplitude. The spins dephasing induced by
the diffusion process is refocused by these control modulations forming
the so-called spin-echoes \citep{Hahn1950,Carr1954} (Fig. \ref{fig:(a)-SDR-sequence}\textbf{a,b}).
At the evolution time $T_{E}$ of the control sequence, the spin-echo
magnetization decays depending on how nuclear spins were scrambled
by the diffusion process. The resulting magnetization at $T_{E}$
is then spatially encoded with an MRI acquisition sequence.

The spins acquire a random phase $\phi(T_{E})$ that typically follows
a Gaussian distribution \citep{Stepisnik1999}. The magnetization
in a voxel of the image becomes\vspace{-2mm}
\begin{equation}
M(T_{E})=e^{-\frac{1}{2}\left\langle \phi^{2}(T_{E})\right\rangle }M(0).\label{eq:M(TE)}
\end{equation}
The quadratic phase is averaged over the spin ensemble as \citep{Stepisnik1993,Grebenkov2007,Callaghan2011}
\vspace{-2mm}

\begin{equation}
\left\langle \phi^{2}(T_{E})\right\rangle \!=\!\gamma^{2}\!\int_{0}^{T_{E}}\!\!\negthinspace\!dt\!\int_{0}^{T_{E}}\!\!\!dt'G(t)G(t')\left\langle \Delta r(t-t')\Delta r(0)\right\rangle ,\label{eq:quadraticdisplacement-1}
\end{equation}
which is expressed in terms of the control applied to the system by
$G(t)$ and the molecular displacement autocorrelation function $\left\langle \Delta r(0)\Delta r(t)\right\rangle =D_{0}\tau_{c}e^{-|t|/\tau_{c}}$
\citep{Stepisnik1993,Alvarez2013a,Shemesh2013}. Here $\Delta r(t)=r(t)-\left\langle r(t)\right\rangle $
is the instantaneous displacement of the spin position from its mean
value, $D_{0}$ is the free diffusion coefficient, and $\tau_{c}$
is the correlation time of the diffusion process. The restriction
length $l_{c}$ of the microstructure compartment in which molecular
diffusion is taking place is then determined by the Einstein-Diffusion
equation $l_{c}^{2}=D_{0}\tau_{c}$ \citep{Callaghan2011}. Then,
by \emph{monitoring the }spin-echo\emph{ decay by applying suitable
control sequences, one can infer microstructure-sizes} \citep{Ong2010,Alvarez2013a,Shemesh2013,Drobnjak2016,Nilsson2017,Novikov2019}.

\textbf{Spin-echo decay-shift as a probing-microstructure paradigm.}
The spin-echo magnetization decay is usually characterized by its
decay-rate \citep{Grebenkov2007,Callaghan2011}. Under the control
sequences discussed in Fig. \ref{fig:(a)-SDR-sequence}\textbf{a,b},
the decay-rate is typically reduced as the number of refocusing periods
$N$ increases \citep{Carr1954}. This effect is shown in Fig. \ref{fig:(a)-SDR-sequence}\textbf{a-c},
where the Hahn spin-echo $(N\!=\!1)$ decay is compared with the signal
after a CPMG sequence ($N\!>\!1$) with multiple echoes \citep{Carr1954}.
However, in the \emph{restricted diffusion} regime when the refocusing
periods are longer than the correlation time $\tau_{c}$, the spin-echo
decays as \textrm{$M(T_{E})/M(0)\approx e^{-\gamma^{2}G^{2}D_{0}\tau_{c}^{2}(T_{E}-(1+2N)\tau_{c})}$}
(see Methods). There, the spins have been fully scrambled within the
compartment, and the dephasing cannot be refocused leading to a decay-rate
$\gamma^{2}G^{2}D_{0}\tau_{c}^{2}$ independent of $N$. Yet, the
spin-echo retains information of the transition from the free to the
restricted diffusion regime. This information is manifested as a ``decay-shift''
$\gamma^{2}G^{2}D_{0}\tau_{c}^{3}(1+2N)$ \emph{independent of the
evolution time} on the spin-echo decay signal, as shown in Fig. \ref{fig:(a)-SDR-sequence}\textbf{c}.
We here demonstrate that this shift can be exploited to selectively
probe microstructure sizes as it is $\propto l_{c}^{6}$.

As this decay-shift depends on $N$\textrm{,} it can be selectively
probed by concatenating a Hahn with a CPMG gradient modulation and
changing the ratio between the relative refocusing periods of each
component of the sequence as shown in Fig. \ref{fig:(a)-SDR-sequence}\textbf{d}.
This control sequence produces an effective modulated gradient $G(t)$,
that conforms the Non-uniform Oscillating-Gradient Spin-Echo (NOGSE)
\citep{Alvarez2013a,Shemesh2013}. This sequence also factorizes out
other relaxation mechanisms allowing to probe selectively the diffusion
induced decay.

We define the NOGSE contrast (NOGSEc) as the amplitude $\Delta M$
given by the difference between the CPMG and the Hahn signal (Fig.
\ref{fig:(a)-SDR-sequence}\textbf{c}). It is obtained by evaluating
NOGSE at the refocusing periods $t_{C}=t_{H}$ and then at $t_{C}\rightarrow0$
using the definitions shown in Fig. \ref{fig:(a)-SDR-sequence}\textbf{d}.
Then both measurements are subtracted (see Methods). Within the restricted
diffusion regime, this contrast amplitude is\vspace{-1mm}

\begin{equation}
\Delta M\approx e^{-\gamma^{2}G^{2}D_{0}\tau_{c}^{3}(\frac{T_{E}}{\tau_{c}}-3)}\!\left(e^{\gamma^{2}G^{2}D_{0}\tau_{c}^{3}2(N-1)}-1\right),\label{eq:NOGSErestr}
\end{equation}
which is very sensitive to the restricted diffusion length $l_{c}$
as it has a parametric dependence $l_{c}^{6}\propto\tau_{c}^{3}$
provided by the spin-echo decay-shift \citep{Alvarez2013a,Shemesh2013}.

\textbf{NOGSE as a selective microstructure-size filter.} NOGSEc \emph{$\Delta M$
has a maximum} as a function of the normalized echo-time $T_{E}/\tau_{c}$
as shown in Fig. \ref{fig:(a)-SDR-sequence}\textbf{e}. We exploit
this maximum contrast to enhance the relative contribution to the
signal from specific restriction lengths $l_{c}$ from a size-distribution.

\begin{figure}
\begin{centering}
\includegraphics[width=1\columnwidth]{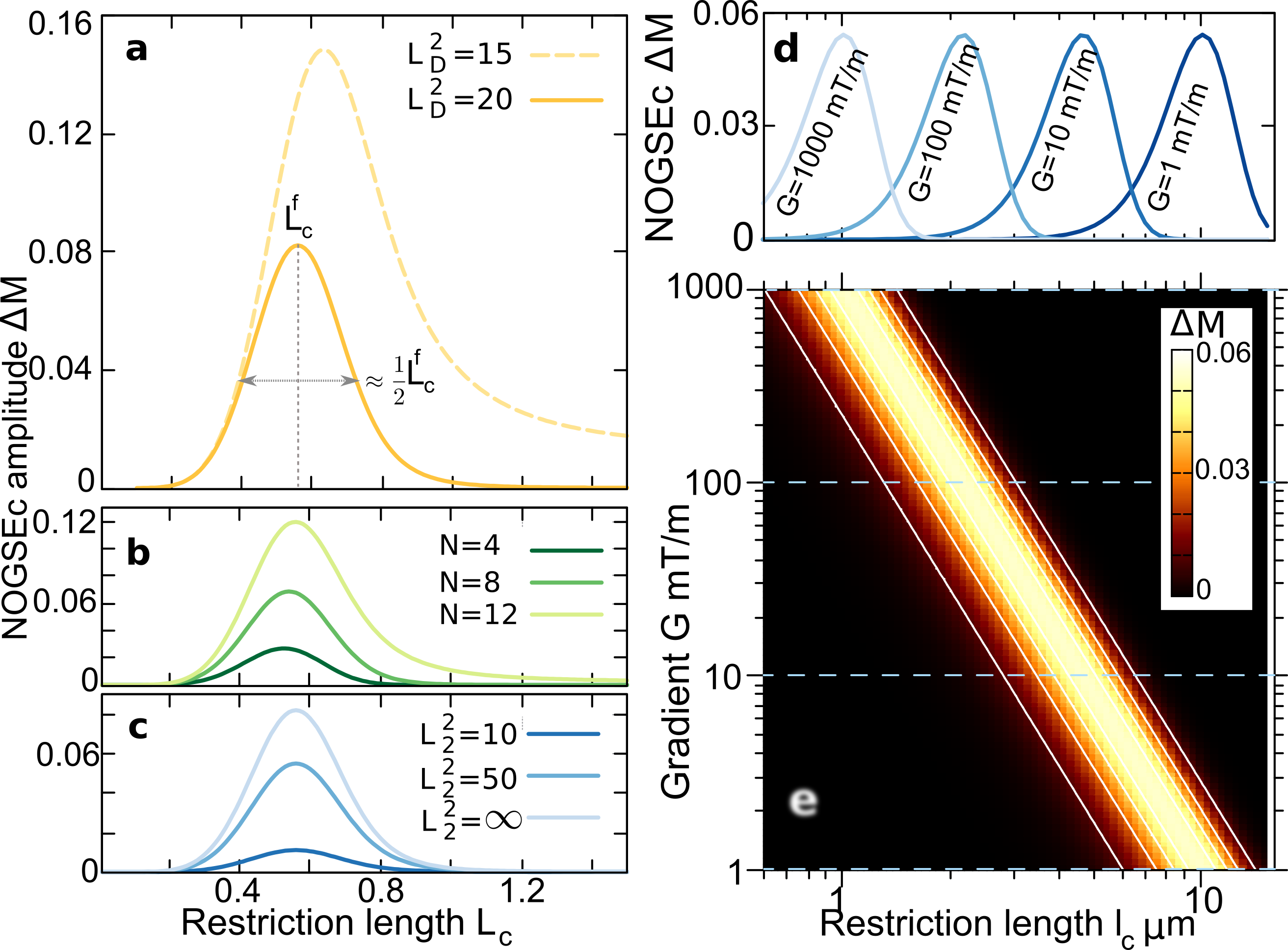}
\par\end{centering}
\caption{\textbf{Selective microstructure-size filter based on NOGSE contrast.}
(\textbf{a}) NOGSEc amplitude \textrm{$\Delta M$} as a function of
the renormalized restriction length $L_{c}$ for different values
of the diffusion length $L_{D}$ ($L_{D}^{2}\!=\!15,20$ with $N\!=\!8$
refocusing periods). \textrm{$\Delta M$} shows a Gaussian filter
functional dependence for $L_{D}\gg1$ with a maximum at \textrm{$L_{c}^{f}\propto L_{D}^{-\frac{1}{2}}$.}
The filter full-width-half-maximun $FWHM\approx\frac{L_{c}^{f}}{2}$.
(\textbf{b}) The amplitude of \textrm{$\Delta M$} at $L_{c}^{f}$
increases linearly with the refocusing periods $N$ as long as $\frac{L_{D}^{2}}{L_{c}^{2}N}\gg1$\textrm{.}
Here \textrm{$L_{D}^{2}=22$.} (\textbf{c})\textrm{ }Attenuation effects
of the filter amplitude due to transversal $T_{2}$-relaxation. Different
dephasing diffusion lengths $L_{2}^{2}\!=\!10,50,\infty$ associated
to $T_{2}$ are shown for $L_{D}^{2}\!=\!20$. $\Delta M$ decreases
with decreasing $L_{2}$. \textbf{(d,e) }\textrm{$\Delta M$} as a
function of the restriction length $l_{c}$ and the gradient strength
\textbf{$G$},\textbf{ }considering $N\!=\!8$. As $L_{D}^{2}\!=\!25$
remains constant by properly varying $T_{E}$, the maximum $\Delta M$
is also constant. The considered dynamic range for the gradient strength
$G$ is achievable with current technologies. The horizontal dashed
lines in (\textbf{e}) correspond to the specific cases plotted in
(\textbf{d}). $L_{c}^{f}$ increases with decreasing $G$. $D_{0}\!=\!0.7\,\mu\mathrm{m^{2}/ms}$
is considered.\label{fig:Filter-TE}}
\end{figure}

In order to perform a general analysis, Eq. (\ref{eq:NOGSErestr})
can be expressed in terms of dimensionless lengths $L_{D}=l_{D}/l_{G},L_{c}=l_{c}/l_{G}$
(see Methods). Here, $l_{c}=\sqrt{D_{0}\tau_{c}},\:l_{D}=\sqrt{D_{0}T_{E}},\:l_{G}=\sqrt[3]{D_{0}/\gamma G}$
are the restriction length, the diffusion length that the spin can
diffuse freely during $T_{E}$ and the dephasing diffusion length
that provides a phase shift of $2\pi$, respectively \citep{Callaghan2011}
(Fig. \ref{fig:(a)-SDR-sequence}\textbf{c}). Then, \textrm{$\Delta M$}
as a function of $L_{c}$ can be approximated by a Gaussian function
when $L_{D}/L_{c}\gg1$, $L_{c}^{6}\ll1$ and $L_{D}\gg1$ (see Methods),\vspace{-1mm}
\begin{equation}
\Delta M\approx2(N-1)e^{-3/2}\left(L_{c}^{f}\right)^{6}\exp\left[-12\left(\frac{L_{c}-L_{c}^{f}}{L_{c}^{f}}\right)^{2}\right].\label{eq:gaussianfilter}
\end{equation}
NOGSEc therefore acts as a microstructure-size ``bandpass-filter''
with $L_{c}^{f}$ as the filter-center size (see Fig. \ref{fig:Filter-TE}\textbf{a})\vspace{-2.5mm}

\begin{equation}
L_{c}^{f}\approx\left(3/2\right)^{\frac{1}{4}}L_{D}^{-\frac{1}{2}}.\label{eq:trascendental}
\end{equation}
The filter-band selectivity is defined by the ratio between the full-width-at-half-maximum
(FWHM) and $L_{c}^{f}$, $FWHM/L_{c}^{f}\approx\sqrt{\frac{\ln2}{3}}\approx0.5$.

The maximum of \textrm{$\Delta M$} at the size $L_{c}^{f}$ can be
tuned to \emph{highlight a given restriction length} $l_{c}$ based
on choosing properly the sequence control parameters, i.e.\textrm{
}the gradient strength $G$ and the evolution time $T_{E}$ (see Fig.
\ref{fig:Filter-TE}\textbf{a}).

The filtered size decreases with increasing the control parameter
$L_{D}$ according to Eq. (\ref{eq:trascendental}). At the same time,
increasing $L_{D}$\textcolor{black}{{} decreases the contrast amplitude}
which is $\propto\left(L_{c}^{f}\right)^{6}\propto1/L_{D}^{3}$. Therefore,
the minimum size that can be filtered in practice is limited by the
Signal-to-Noise Ratio (SNR) and the maximum achievable gradient strength
as $L_{D}\propto\sqrt[3]{G}$. This decrease of $\Delta M$ can be
compensated linearly with increasing the number of refocusing periods
$N$, as long as $\frac{L_{D}^{2}}{L_{c}^{2}N}=\frac{T_{E}}{\tau_{c}N}\gg1$
to reach the restricted regime (see Eq. (\ref{eq:gaussianfilter})
and Fig. \ref{fig:Filter-TE}\textbf{b}).

A practical limitation is also the unavoidable transversal $T_{2}$-relaxation
due to the intrinsic dephasing of the nuclear spins. NOGSEc decays
then as\vspace{-2mm}
\begin{equation}
\Delta M_{T_{2}}=\Delta M\,e^{-L_{D}^{2}/L_{2}^{2}},\label{eq:DeltaM_Nogse(T2)}
\end{equation}
where we have defined \textrm{$L_{2}=\frac{l_{2}}{l_{G}}$} as the
$T_{2}$ diffusion dimensionless length with $l_{2}=(D_{0}T_{2})^{1/2}$.
The $T_{2}$-relaxation effect is showed in Fig. \ref{fig:Filter-TE}\textbf{c}
for different values of $L_{2}^{2}$. Remarkably, the filter shape
and center remains the same as the case of $T_{2}\rightarrow\infty$.

\begin{figure}
\begin{centering}
\includegraphics[width=1\columnwidth]{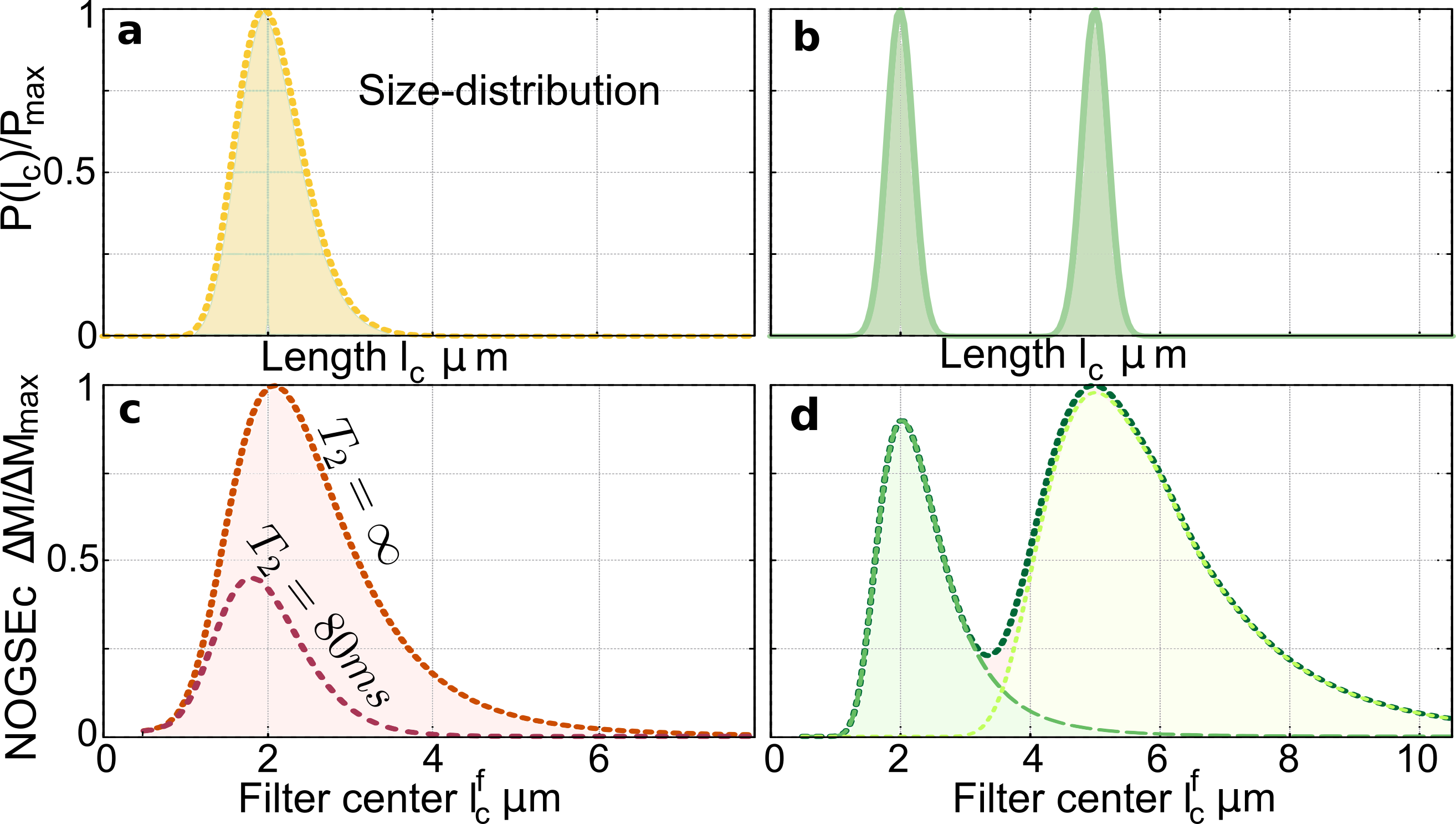}
\par\end{centering}
\caption{\textbf{Selective size-filtering with NOGSE contrast in typical microstructure
size-distributions.} Size-distribution probability \textrm{$P(l_{c})$}
for two typical cases of tissue microstructure, (\textbf{a}) a log-normal
distribution with median $2\,\mathrm{\mu m}$ and geometric standard
deviation $1.22\,\mathrm{\mu m}$, representative of white matter
size-distributions; and (\textbf{b}) a bimodal Gaussian distribution
with means $l_{c}\!=\!2\,\mathrm{\mu m}$ and $l_{c}\!=\!5\,\mathrm{\mu m}$,
and a standard deviation $0.2$$\mathrm{\mu m}$ for both peaks. (\textbf{c,d})
Normalized NOGSEc amplitude $\Delta M$ as a function of the filter
center $l_{c}^{f}$. The diffusion coefficient is $D_{0}\!=\!0.7\,\mu\mathrm{m^{2}/ms}$
in both cases. \textbf{(c)} $\Delta M$ including transversal $T_{2}$-relaxation
effects (dashed line, $T_{2}\!=\!80\,\mathrm{ms}$) is contrasted
with the ideal case without relaxation effects (solid line, $T_{2}\!=\!\infty$)
for the distribution of panel (\textbf{a}). Here $L_{D}^{2}\!=\!11$,
$N\!=\!4$ and $\Delta M_{max}\!=\!0.08$. \textbf{(d)} The dashed
lines show $\Delta M$ obtained considering separately each of the
component of the bimodal distribution of panel (\textbf{b}). The low
overlap between the dashed lines demonstrates the filtering property
of NOGSEc. Here $L_{D}^{2}\!=\!25$, $N\!=\!8$ and $\Delta M_{max}\!=\!0.027$.\label{filtersweep}}
\end{figure}

\textbf{Selective size-filtering in typical microstructure-size distributions.
}Our results show that one can produce quantitative images based
on a signal contrast generated by specific microstructure-sizes from
a size-distribution. This method avoids extracting the microstructure-size
by fitting a curve, which is time consuming as it typically requires
several measurements \citep{Assaf2008,Alexander2010,Ong2010,Xu2014,Shemesh2015,Novikov2019}.
The filter amplitude $\Delta M$ in Eq. (\ref{eq:gaussianfilter})
remains constant by varying the gradient strength and the evolution
time keeping fixed the parameter $L_{D}$, as shown in Fig. \ref{fig:Filter-TE}\textbf{d-e}.
We exploit this property for a proof-of-principle evaluation of the
performance of the NOGSEc filter applied to microstructural size-distribution
inherent to heterogeneous biological tissues as shown in Fig. \ref{filtersweep}.
Typical size-distributions $P(l_{c})$ are log-normal and bi-modal
functions \citep{Assaf2008,Pajevic2013,White2013,Liewald2014,Shemesh2015}
as shown in Figs. \ref{filtersweep}\textbf{a-b}. NOGSEc for a size-distribution
is given by 
\begin{equation}
\Delta M=\int_{l_{c}}P(l_{c})\Delta M(l_{c})dl_{c},\label{eq:Distribution}
\end{equation}
 where $\Delta M(l_{c})$ is the contribution for a given restriction
length $l_{c}$. Figures \ref{filtersweep}\textbf{c-d} show $\Delta M$
as a function of $l_{c}^{f}=L_{c}^{f}l_{G}$, where $G$ and $T_{E}$
are changed simultaneously keeping $L_{D}^{2}$ constant. The center
of the filter at $l_{c}^{f}$ is therefore swept while the filter
amplitude is kept constant. The resulting filtered signal as a function
of the filter center $l_{c}^{f}$ therefore resembles the original
size-distributions, where the log-normal peak and both Gaussian peaks
are clearly identified. The effects of $T_{2}$-relaxation are shown
in Fig. \ref{filtersweep}\textbf{c} for typical values of white-matter
tissue. To further demonstrate the filter selectivity, Figure \ref{filtersweep}\textbf{d}
shows that if one chooses $l_{c}^{f}$ equal to the center of one
of the two Gaussian distributions, the other component is filtered-out.
These simulations demonstrate the feasibility of performing \emph{quantitative
images of a selective microstructure-size $l_{c}^{f}$ based on the
NOGSEc amplitude as a ``bandpass'' filter}.

\begin{figure}
\begin{centering}
\includegraphics[width=1\columnwidth]{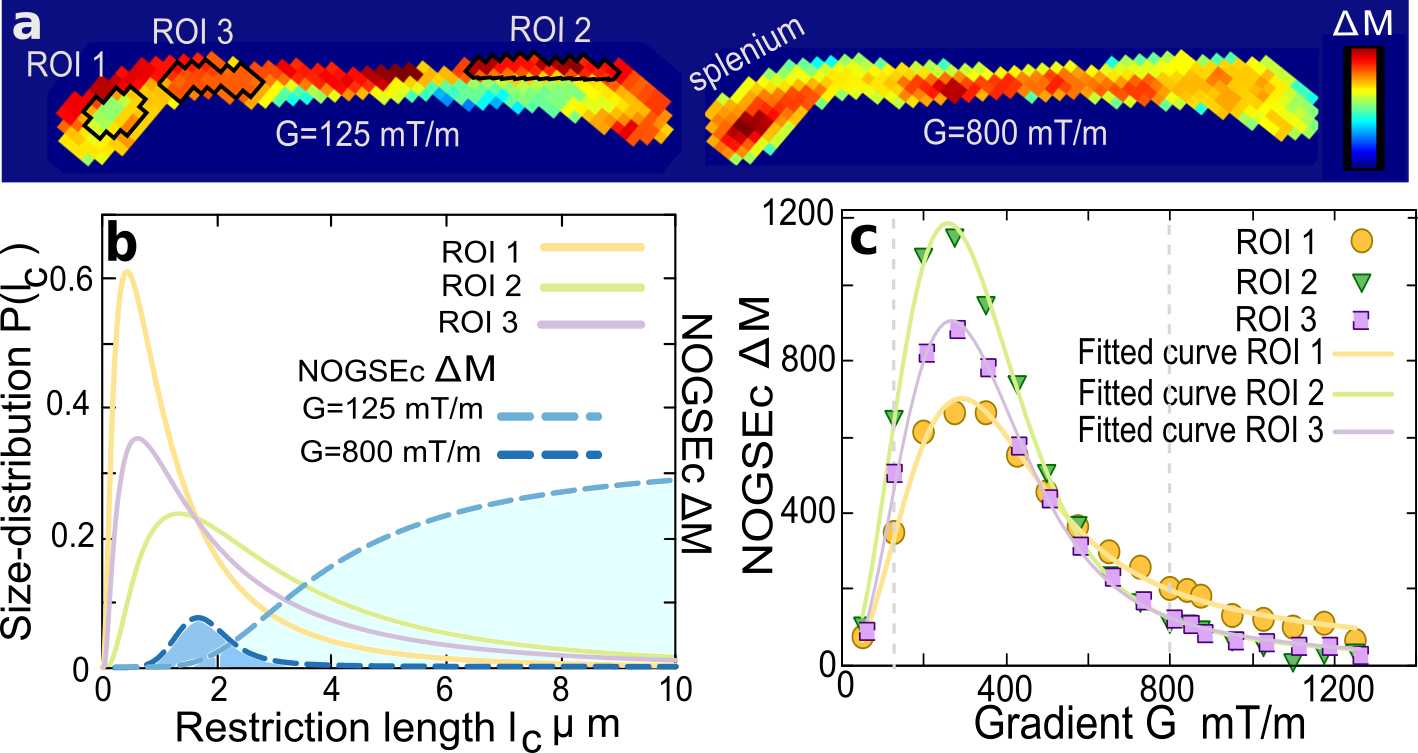}
\par\end{centering}
\caption{\textbf{Non-invasive NOGSE-imaging of selective microstructure-sizes
in ex-vivo mouse brain.} (\textbf{a}) Two images based on NOGSEc of
the Corpus Callosum\textbf{ }region of a mouse brain for two gradient
strengths. The $\Delta M$ contrast highlights different zones with
complementary colors when comparing the images for the two gradients.
The color scale covers the full range of contrast signal in each image.
Three regions-of-interest (ROI) are indicated in black contours. The
images were acquired for $N\!=\!2$ and $T_{E}\!=\!21.5\,\mathrm{ms}$.
Pixel-size $78\times78\mu\text{m}^{2}$. (see more details in Methods).
\textbf{(b) }Size-distributions (solid-lines) that best fit the experimental
data for the selected ROIs. The dashed lines show $\Delta M(l_{c})$
predicted by our model for the two gradients used in (\textbf{a}).
The predicted overlap between the reconstructed distribution and the
$\Delta M(l_{c})$ filter is consistent with the signal contrast shown
in (\textbf{a} and \textbf{c}): ROI 1 has lower sizes than ROI 2 and
3, and therefore ROI 1 has higher $\Delta M$ for $G\!=\!800\,\mathrm{mT/m}$;
conversely ROI 2 and 3 have higher contrast amplitude for $G\!=\!125\,\mathrm{mT/m}$.(\textbf{c})
Average NOGSEc signal (symbols) as a function of the gradient strength
$G$ for the three ROIs. The vertical dashed lines mark the gradient
strengths used in (\textbf{a}). The solid lines are fits to the experimental
data of our theoretical model following Eq. (\ref{eq:Distribution})
for a log-normal distribution. The fitted parameters are the median
$1.08\pm0.06$, $2.82\pm0.01$ and $1.87\pm0.04\,\mu m$ and the geometric
standard deviation $2.58\pm0.03$, $2.39\pm0.02$ and $2.91\pm0.04\,\mu m$
for ROI 1, 2 and 3, respectively.\textbf{ }We considered a uniform
$D_{0}\!=\!0.7\,\mu\mathrm{m^{2}/ms}$ as a representative diffusion
coefficient to fit the size-distribution.\label{fig:Experimental}\textrm{}}
\end{figure}

The transversal $T_{2}$-relaxation limits the largest microstructure-size
that can be filtered, as the restricted diffusion regime has to be
achieved. A good SNR for $\Delta M$ is only obtained for $L_{c}<L_{D}<L_{2}$.
For short $T_{2}$, the strategy is therefore using the lowest possible
value of $L_{D}$ and reducing the number of refocusing periods $N$
so as to remain in the restricted regime. We consider this scenario
in proof-of-principle experiments and perform a size-filter sweep
varying only the gradient amplitude while keeping $T_{E}\lesssim T_{2}$
constant. We implement the microstructure-size filter method on an
ex-vivo mouse brain focusing on the Corpus Callosum (CC) region (see
experimental details in Methods). The CC contains aligned axons and
is a paradigmatic model for log-normal size-distributions \citep{Assaf2008,Pajevic2013,White2013,Liewald2014,Shemesh2015}.
Figure \ref{fig:Experimental}\textbf{a} shows images of $\Delta M$
for two gradient strengths. The largest gradient acts as a ``bandpass''
Gaussian-filter of the lower microscopic sizes of the distribution,
compared to the weaker gradient that acts as a ``high-pass'' filter
of the larger sizes (Fig. \ref{fig:Experimental}\textbf{b}). Therefore
Fig. \ref{fig:Experimental}\textbf{a} clearly highlights zones of
the CC with complementary colors depending of the microstructure-size.
This is demonstrated quantitatively in Fig. \ref{fig:Experimental}\textbf{b-c}
for three regions-of-interest (ROI). The average \textrm{$\Delta M$}
as a function of the gradient for the ROIs is shown in Fig. \ref{fig:Experimental}\textbf{c}
together with fitted curves derived from our theoretical model following
Eq. (\ref{eq:Distribution}) (see Methods). The inferred microstructure-size
distributions and the $\Delta M$ filter shapes are shown in Fig.
\ref{fig:Experimental}\textbf{b}. The excellent agreement between
the model and the experimental data fully demonstrates the reliability
of the quantitative images shown in panel \textbf{a} based on the
NOGSE microstructure-size filter and the assumed log-normal model.\textrm{}

\textbf{Conclusions.} The presented results introduce a method for
performing non-invasive quantitative images of selective microstructure-sizes
based on probing nuclear-spin dephasing induced by molecular diffusion
with magnetic resonance. Conversely to standard diffusion-weighted
imaging approaches that are based on observing the decay-rate of the
spin signal, we exploit dynamical control with oscillating gradients
to selectively probe a decay-shift on spin-echo decays. This decay-shift
contains quantitative information of microstructure-sizes that restrict
the molecular diffusion. We generate a contrast amplitude that behaves
as a microstructure-size filter to selectively probe a restriction
length determined by the control parameters. We show the usefulness
and performance of the method with proof-of-principle simulations
and experiments on typical size-distributions of white-matter tracts
in a mouse brain. A quantitative image of specific diffusion restriction
lengths is performed extracted from only two images, allowing to significantly
reduce the tens of images that typically demand the inference of microstructure-sizes
from data fittings \citep{Assaf2008,Alexander2010,Ong2010,Xu2014,Shemesh2015,Novikov2019}.
Even though intrinsic $T_{2}$-relaxation may represent a limitation,
we show excellent performance probing quantitative information of
microstructure details between $\sim0.1-10\,\mu\mathrm{m}$ on biological
tissue as in the mouse white matter, being able to filter sizes much
lower than the present image resolution. This work lays the foundations
of a novel conceptual tool with low overhead for designing quantitative
methods for non-invasive imaging of tissue microstructure. This diagnostic
tool opens up a new avenue to explore for in-vivo imaging. In addition,
these results can also be applied for characterizing material microstructures,
such as rocks which are of particular interest for oil extraction,
and for nanoscale-imaging of biological tissues with novel quantum
sensors based on noise spectroscopy \citep{Staudacher2015,Barry14133,Wangeaau8038}.
\begin{acknowledgments}
\textbf{Acknowledgments.} We acknowledge Soledad Esposito and Micaela
Kortsarz for preparing the ex-vivo mouse brain, and Federico Turco
for scripting assistance to process the experimental data. We thank
Lucio Frydman and Jorge Jovicich for fruitful discussions. This work
was supported by CNEA, ANPCyT-FONCyT PICT-2017-3447, PICT-2017-3699,
PICT-2018-04333, PIP-CONICET (11220170100486CO), UNCUYO SIIP Tipo
I 2019-C028, Instituto Balseiro. A.Z. and G.A.A. are members of the
Research Career of CONICET. M.C. and P.J. acknowledge support from
the Instituto Balseiro's fellowships.
\end{acknowledgments}

\bibliographystyle{apsrev4-1}
\bibliography{bibliography,bibliography-press}

\newpage{}

\clearpage{}

\section*{Methods}

\textbf{Magnetization decay of an spin ensemble under dynamical control.
}The magnetization signal observed from an ensemble of non-interacting
and equivalent spins, under the effect of dynamical control, is $M(t)=\left\langle e^{-i\phi(t)}\right\rangle M(0)$.
Here, the brackets denote the ensemble average over the random phases
$\phi(t)$ acquired by the spins during the evolution time $t$. For
the considered dynamical control with modulated gradient spin-echo
sequences, the average phase becomes null, $\left\langle \phi(t)\right\rangle =0$.
Then, as $\phi(t)$ typically follows a Gaussian distribution \citep{Stepisnik1999},
the signal will depend on the random phase variance $M(t)=e^{-\frac{1}{2}\left\langle \phi^{2}(t)\right\rangle }M(0)$.

The variance expressed in terms of the control applied to the system
by $G(t)$ and the molecular displacement autocorrelation function
$\left\langle \Delta r(0)\Delta r(t)\right\rangle =D_{0}\tau_{c}e^{-|t|/\tau_{c}}$
\citep{Stepisnik1993,Alvarez2013a,Shemesh2013} is given in Eq. (\ref{eq:quadraticdisplacement-1})
of the main text.

For a piecewise constant modulation $G(t)$, that switches $N$ times
its sign at times $t_{i}$ with $i=0..N-1$ during the evolution time
$T_{E}$. The quadratic phase of the magnetization decay is

\begin{widetext}

\begin{align}
\left\langle \phi^{2}(T_{E})\right\rangle  & ={\displaystyle \gamma^{2}G^{2}D_{0}^{2}\tau_{c}\stackrel[i=0]{N-1}{\sum}\stackrel[j=0]{N-1}{\sum}\int_{t_{i}}^{t_{i+1}}\int_{t_{j}}^{t_{j+1}}}e^{-|t-t'|/\tau_{c}}(-1)^{i}(-1)^{j}{\rm d}t'\,{\rm d}t\nonumber \\
 & {\displaystyle =\gamma^{2}G^{2}D_{0}^{2}\tau_{c}\stackrel[i=0]{N-1}{\sum}\stackrel[j=0]{N-1}{\sum}}(-1)^{i}(-1)^{j}\left[\left(\begin{cases}
2\tau_{c}t_{j}-\tau_{c}^{2}e^{-\frac{t_{i}-t_{j}}{\tau_{c}}} & t_{j}\leq t_{i}\\
-\tau_{c}^{2}e^{-\frac{t_{i}-t_{j}}{\tau_{c}}} & t_{i}<t_{j}
\end{cases}\right)\right.\nonumber \\
 & +\left(\begin{cases}
2\tau_{c}t_{j+1}-\tau_{c}^{2}e^{-\frac{t_{i+1}-t_{j+1}}{\tau_{c}}} & t_{j+1}\leq t_{i+1}\\
-\tau_{c}^{2}e^{\frac{t_{i+1}-t_{j+1}}{\tau_{c}}}+2\tau_{c}t_{i+1} & t_{i+1}<t_{j+1}
\end{cases}\right)-\left(\begin{cases}
2\tau_{c}t_{j}-\tau_{c}^{2}e^{-\frac{t_{i+1}-t_{j}}{\tau_{c}}} & t_{j}\leq t_{i+1}\\
-\tau_{c}^{2}e^{-\frac{t_{i+1}-t_{j}}{\tau_{c}}}+2\tau_{c}t_{i+1} & t_{i+1}<t_{j}
\end{cases}\right)\label{eq:general-decay}\\
 & \left.-\left(\begin{cases}
2\tau_{c}t_{j+1}-\tau_{c}^{2}e^{-\frac{t_{i}-t_{j+1}}{\tau_{c}}} & t_{j+1}\leq t_{i}\\
-\tau_{c}^{2}e^{\frac{t_{i}-t_{j+1}}{\tau_{c}}}+2\tau_{c}t_{i} & t_{i}<t_{j+1}
\end{cases}\right)\right].\nonumber 
\end{align}

\end{widetext}

\textbf{Magnetization decay within the restricted diffusion regime.
}In the restricted diffusion regime all terms $\propto e^{-\frac{\left|t_{i}-t_{j}\right|}{\tau_{c}}}\rightarrow0$
as \textrm{$\left|t_{i}-t_{j}\right|\gg\tau_{c}$} for all $i\ne j$,
therefore the non-null terms in Eq. (\ref{eq:general-decay}) are
those with $i=j$. The phase variance is then

\begin{widetext}

\begin{align}
-\frac{1}{2}\left\langle \phi^{2}(T_{E})\right\rangle  & {\displaystyle =-\frac{1}{2}{\displaystyle \gamma^{2}G^{2}D_{0}^{2}\tau_{c}\stackrel[i=0]{N-1}{\sum}\stackrel[j=0]{N-1}{\sum}}(-1)^{i}(-1)^{j}}\left[\left(2\tau_{c}t_{j}-\tau_{c}^{2}e^{-\frac{t_{i}-t_{j}}{\tau_{c}}}\right)\delta_{ij}+\left(2\tau_{c}t_{j+1}-\tau_{c}^{2}e^{-\frac{t_{i+1}-t_{j+1}}{\tau_{c}}}\right)\delta_{i+1j+1}\right.\\
 & \:\left.-\left(2\tau_{c}t_{j}-\tau_{c}^{2}e^{-\frac{t_{i+1}-t_{j}}{\tau_{c}}}\right)\delta_{i+1j}-\left(2\tau_{c}t_{j+1}-\tau_{c}^{2}e^{-\frac{t_{i}-t_{j+1}}{\tau_{c}}}\right)\delta_{ij+1}\right]\nonumber \\
 & {\displaystyle =-\frac{1}{2}{\displaystyle \gamma^{2}G^{2}D_{0}^{2}\tau_{c}}}\left[2\tau_{c}\stackrel[i=1]{N}{\sum}\left(t_{i}-t_{i-1}\right)-\tau_{c}^{2}(4N+2)\right]\\
 & {\displaystyle =-\gamma^{2}G^{2}D_{0}^{2}\tau_{c}^{2}}\left[T_{E}-(2N+1)\tau_{c}\right].
\end{align}

\end{widetext}

Then, the \emph{decay-rate} is $\frac{d\left\langle \phi^{2}(T_{E})\right\rangle }{dT_{E}}=\gamma^{2}G^{2}D_{0}^{2}\tau_{c}^{2}$,
and the \emph{decay-shift} is the time-independent term ${\displaystyle \gamma^{2}G^{2}D_{0}^{2}\tau_{c}^{3}}(2N+1)$,
which can also be derived from

\begin{equation}
\gamma^{2}G^{2}D_{0}\tau_{c}^{3}(2N+1)\!=\!\gamma^{2}G^{2}D_{0}\tau_{c}(2N+1)\!\!\int_{0}^{\infty}\!\!\!\!dt\,t\left\langle \Delta r(0)\Delta r(t)\right\rangle .\label{eq:Shift-1}
\end{equation}

\textbf{NOGSE contrast amplitude. }One can obtain an analytical expression
for the magnetization decay in Eq. (\ref{eq:general-decay}) for the
Hahn, CPMG and NOGSE spin-echo sequences described in Fig. (\ref{fig:(a)-SDR-sequence})
of the main text. In the restricted diffusion regime $T_{E},t_{\mathrm{H}},t_{\mathrm{C}}\gg\tau_{c}$,
they result

\begin{equation}
\begin{split} & M_{\mathrm{Hahn}}(t_{\mathrm{H}})=\mathrm{exp}\{-\gamma^{2}G^{2}D_{0}\tau_{c}^{3}[\frac{t_{\mathrm{H}}}{\tau_{c}}-3]\},\\
 & M_{\mathrm{CPMG}}(Nt_{\mathrm{C}},N)=\mathrm{exp}\{-\gamma^{2}G^{2}D_{0}\tau_{c}^{3}[\frac{Nt_{\mathrm{C}}}{\tau_{c}}-(2N+1)]\},\\
 & M_{NOGSE}(T_{E},N,t_{C})=\mathrm{exp}\{-\gamma^{2}G^{2}D_{0}\tau_{c}^{3}[\frac{T_{E}}{\tau_{c}}-(2N+1)]\},
\end{split}
\label{eq:Mcpmg-Mhahn}
\end{equation}
with $T_{E}=(N-1)t_{\mathrm{C}}+t_{H}$.

We define the NOGSE contrast (NOGSEc) amplitude $\Delta M$ to the
difference between \textrm{$M_{NOGSE}(T_{E},N,t_{C}=t_{H})=M_{\mathrm{CPMG}}(\frac{T_{E}}{N},N)$
and $M_{NOGSE}(T_{E},N,t_{C}\rightarrow0)\simeq M_{\mathrm{Hahn}}(T_{E})$}),
i.e. 
\begin{equation}
\Delta M(T_{E},N)=\mathrm{M_{\mathrm{CPMG}}(T_{E},N)-M_{\mathrm{Hahn}}(T_{E})}.\label{eq:NogseComplete}
\end{equation}
Then, we arrive to Eq. (\ref{eq:NOGSErestr}) of the main text by
introducing Eq. (\ref{eq:Mcpmg-Mhahn}) into Eq. (\ref{eq:NogseComplete})

\begin{equation}
\Delta M\approx e^{-\gamma^{2}G^{2}D_{0}\tau_{c}^{3}(\frac{T_{E}}{\tau_{c}}-3)}\!\left(e^{\gamma^{2}G^{2}D_{0}\tau_{c}^{3}2(N-1)}-1\right).
\end{equation}
NOGSEc results

\begin{equation}
\Delta M(L_{D},L_{c},N)=e^{-L_{c}^{4}(L_{D}^{2}-3L_{c}^{2})}(e^{2(N-1)L_{c}^{6}}-1),\label{eq:NogseContRest}
\end{equation}
within the restricted diffusion regime, using the dimensionless variables
$L_{D},L_{c}$ defined in the main text.

The general expression for $\Delta M$ that includes all diffusion
time scales can be obtained from Eq. (\ref{eq:general-decay}) by
replacing the time intervals as defined in Fig. \ref{fig:(a)-SDR-sequence}\textbf{c}
of the main text.

\textbf{Gaussian microstructure-size filter derivation. }The NOGSEc
amplitude in the restricted diffusion regime, Eq. (\ref{eq:NogseContRest}),
can be approximated by 
\begin{equation}
\Delta M\approx e^{-L_{c}^{4}(L_{D}^{2}-3L_{c}^{2})}2(N-1)L_{c}^{6},\label{eq:DeltaM_taylor}
\end{equation}
for $L_{c}\ll1$. The maximum of $\Delta M$ occurs at $\frac{d\Delta M}{dL_{C}}=0$.
In the asymptotic limit of $L_{D}\gg1$, it is achieved for 
\begin{equation}
L_{c}=L_{c}^{f}\approx\left(3/2\right)^{\frac{1}{4}}L_{D}^{-\frac{1}{2}}+\mathcal{O}(L_{D}^{-\frac{7}{2}}),
\end{equation}
where \textrm{$L_{c}^{f}$} is the center of the filter as described
in Eq. (\ref{eq:trascendental}) of the main text.

The NOGSEc amplitude can be approximated by

\begin{eqnarray}
\Delta M & \approx & {\rm e}^{-3/2}3\sqrt{\frac{3}{2}}\left(N-1\right)L_{D}^{-3}\\
 &  & -36{\rm e}^{-3/2}\left(N-1\right)L_{D}^{-2}(L_{c}-L_{c}^{f})^{2}\!+\!\mathcal{O}((L_{c}-L_{c}^{f})^{4}),\nonumber 
\end{eqnarray}
with a Taylor expansion in $L_{c}$ at $L_{c}\approx L_{c}^{f}$ of
the expression given in Eq. (\ref{eq:DeltaM_taylor}). We use this
expansion to define the first moments of the Gaussian filter function
of Eq. (\ref{eq:gaussianfilter}) in the main text, obtaining

\begin{equation}
\Delta M\approx2(N-1)e^{-3/2}\left(L_{c}^{f}\right)^{6}\exp\left[-12\left(\frac{L_{c}-L_{c}^{f}}{L_{c}^{f}}\right)^{2}\right].\label{eq:gaussianfilter-1}
\end{equation}
This expression is then verified to approximate very well the exact
expression derived from Eq. (\ref{eq:general-decay}) within the regime
of $L_{D}\gg1$ and $L_{c}\ll1$.

\textbf{Ex-vivo mouse brain preparation. }The experiments were approved
by the Institutional Animal Care and Use Committee of the Comisión
Nacional de Energía Atómica under protocol number 08\_2018. One mouse
was sacrificed by isoflurane overdose and its brain was fixed in formaline.
The brain was washed twice with PBS prior to the insertion into a
15 ml falcon tube filled with PBS. The brain was left in the magnet
for at least three hours prior to the reported experiments to reach
thermal equilibration.

\begin{figure}
\begin{centering}
\includegraphics[width=1\columnwidth]{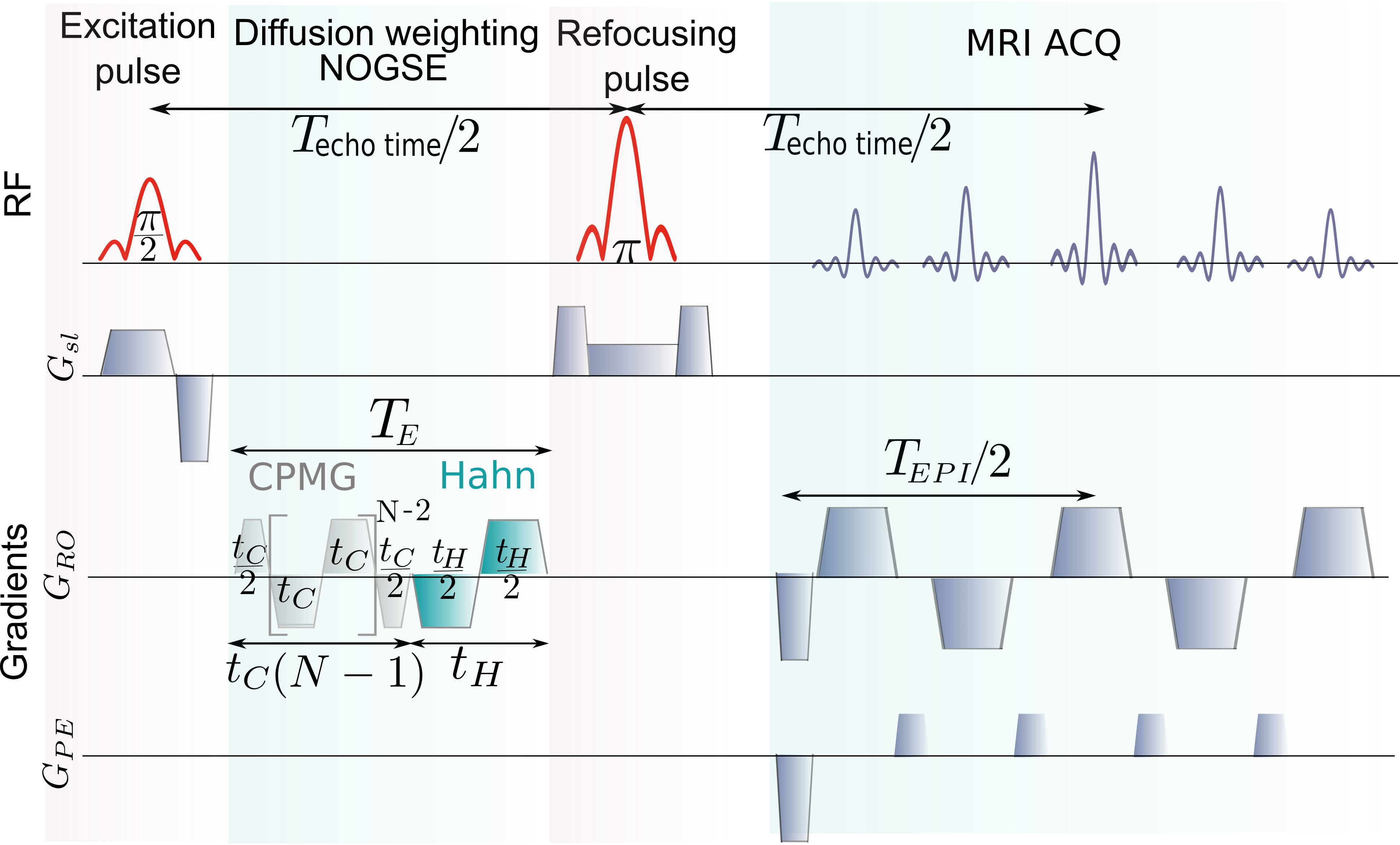}
\par\end{centering}
\caption{\textbf{\label{fig:NOGSEexpSequence}Scheme for the experimental implementation
of the NOGSE sequence.} An initial selective RF-excitation $\frac{\pi}{2}$-pulse
is applied to select a tissue slice. It is followed by a NOGSE gradient
modulation of duration $T_{E}$ following the scheme described in
Fig. \ref{fig:(a)-SDR-sequence}\textbf{d} of the main text. During
the evolution time $T_{E}$, the gradient strength and sign is modulated
with trapezoidal shapes. Then a selective RF $\pi$-pulse is applied
to refocus magnetic field inhomogeneities. At the end a spatial EPI-enconding
is applied for acquiring an image. Three gradients are applied in
the three spatial directions for slide selection $G_{sl}$, for read
orientation $G_{RO}$ and phase encoding $G_{PE}$. The NOGSE gradients
can be applied in arbitrary orientations.}
\end{figure}

\textbf{MRI experiments. }The experiments were performed on a 9.4T
Bruker Avance III HD WB NMR spectrometer with a 1H resonance frequency
of $\omega_{z}=400.15$ MHz. We use a Micro 2.5 probe capable of producing
gradients up to 1500 mT/m in three spatial directions. The experiments
temperature was stabilized at 21°C. We programmed and implemented
with Paravision 6 the NOGSE MRI sequence shown in Fig. \ref{fig:NOGSEexpSequence}.
The sequence parameters were: Repetition time $2000$ ms, $T_{\text{echo time}}=55\text{ms}$,
FOV = 15x15 mm$^{2}$ with a matrix size of 192x192, leading to an
in-plane resolution of $78\times78\mu\text{m}^{2}$, and slice thickness
of 1 mm with 128 signal averages. The two images were acquired with
echo planar imaging (EPI) encoding with 4 segments (image acquisition
time $\approx17$ min) and then subtracted to generate $\Delta M$.
The NOGSE modulation time was $T_{E}=21.5\text{ms}$ with $N=2$.
The NOGSE gradients were applied perpendicular to the main axis of
the axons in the corpus callosum. NOGSEc $\Delta M$ is determined
from an image generated with $t_{H}=t_{C}=10.75\text{ms}$ for the
CPMG modulation and with $t_{H}=0.5\text{ms}$ and $t_{C}=21\text{ms}$
for the Hahn modulation. The set of parameter values were chosen for
achieving good SNR for performing the proof-of-principle experiments.
Further studies should be considered to explore the optimal values
for acquiring the images in the shortest possible time.

\textbf{Experimental data analysis. }The mean signal from the pixels
in the ROIs of Fig. \ref{fig:Experimental}\textbf{a} of the main
text was analyzed, and plotted as a function of $G$ in Fig. \ref{fig:Experimental}\textbf{c}.
Fittings to the theoretical model were done assuming a uniform $D_{0}=0.7\mu m^{2}/\text{ms}$
and a log-normal distribution $P(l_{c})=\frac{1}{\sqrt{2\pi}\sigma l_{c}}e^{-\frac{(ln(l_{c})-\mu)^{2}}{2\sigma^{2}}}$
with median $e^{\mu}$ and geometric standard deviation $e^{\sigma}$.
This implies that no extra assumptions were considered for the tissue
model (e.g., intra/extra-cellular compartments). Therefore a single
log-normal distribution was thus fitted to the experimental data,
regardless of the potential heterogeneity. This means that all underlying
compartments (e.g., extracellular, intracellular, etc.) reflected
in the diffusion weighted are assumed to be described by a single
log-normal distribution. We considered a distribution of restriction
lengths $l_{c}$ without assuming particular geometries. Remarkably
the excellent agreement of the fitted curves to the experimental data
in Fig. \ref{fig:Experimental}\textbf{c} is consistent with these
simple assumptions.
\end{document}